\documentclass[showpacs,pre,aps,twocolumn,superscriptaddress]{revtex4-1}

\pdfoutput=1
\usepackage{amsmath,amsfonts,amssymb,bm,graphicx,color}

\usepackage{hyperref}
\usepackage{color}

\usepackage{graphicx}
\usepackage{dcolumn}
\usepackage{bm}
\usepackage{amsmath}
\usepackage{setspace}
\usepackage{color}
\usepackage{float}
\usepackage{subfigure}
\usepackage{array}
\usepackage{xr}
\usepackage[normalem]{ulem}

\begin{document}

\title{Study of the upper-critical dimension of the East model through the breakdown of the Stokes-Einstein relation}

\author{Soree Kim}
\altaffiliation{Equal contribution}
\affiliation{Department of Chemistry, Seoul National University, Seoul 08826, Republic of Korea}

\author{Dayton G. Thorpe}
\altaffiliation{Equal contribution}
\affiliation{Department of Chemistry, University of California, Berkeley, California 94720, USA}

\author{Juan P. Garrahan}
\affiliation{School of Physics and Astronomy, University of Nottingham, Nottingham, NG7 2RD, UK}

\author{David Chandler}
\affiliation{Department of Chemistry, University of California, Berkeley, California 94720, USA}

\author{YounJoon Jung\thanks{}}%
\email[Electronic mail:]{yjjung@snu.ac.kr}
\affiliation{Department of Chemistry, Seoul National University, Seoul 08826, Republic of Korea}

\date{\today}

\begin{abstract}
We investigate the dimensional dependence of 
dynamical fluctuations related to 
dynamic heterogeneity in supercooled liquid systems 
using kinetically constrained models. 
The $d$-dimensional spin-facilitated East model with embedded
probe particles is used as a representative super-Arrhenius glass forming system. 
We investigate the existence of an upper critical dimension in this model
by considering decoupling of transport 
rates through an effective fractional Stokes-Einstein relation, $D\sim{\tau}^{-1+\omega}$, with $D$ and $\tau$ the diffusion constant of the probe particle
and the relaxation time of the model liquid, respectively, and where $\omega > 0$ encodes the breakdown of the standard Stokes-Einstein relation. 
To the extent that decoupling indicates non mean-field behavior, our simulations suggest that the East model has an upper critical dimension which is at least above $d=10$, and argue that it may be actually be infinite. 
This result is due to the existence of hierarchical dynamics in the East model in any finite dimension.  We discuss the relevance of these results for studies of decoupling in high dimensional atomistic models.
\end{abstract}

\maketitle

\section{Introduction}
The East model and its higher dimensional generalizations 
\cite{Jackle1991,Ritort2003,Berthier2005,Ashton2005} 
describes the cooperative relaxation dynamics of glass formers through a simple facilitation mechanism.  This simple model captures fundamental features of the dynamics close to the glass transition, such as super-Arrhenius growth of the relaxation time \cite{Sollich1999}, dynamic heterogeneity \cite{Garrahan2002}, transport decoupling \cite{Jung2004,Jung2005,Berthier2005E,Blondel2014}, the existence of space-time transitions \cite{Merolle2005,Garrahan2007}, thermodynamic anomalies under cooling \cite{Keys2013} and melting of ultrastable glasses \cite{Gutierrez2016}.  (For reviews on the glass transition problem see for example \cite{Ediger1996,Berthier2011,Biroli2013}).  

The theoretical perspective on the glass transition that emerges from the study of the East model and other kinetically constrained models (KCMs) - sometimes called dynamic facilitation (DF) theory - is one of fluctuation dominance in the dynamics with a very limited role played by the thermodynamics of glass formers (see \cite{Chandler2010} for a review).  This contrasts with theoretical approaches based on mean-field theory, in particular that of the random first-order transition (RFOT) perspective (see \cite{Lubchenko2007,Parisi2010} for reviews).  Within RFOT, mean-field becomes exact above an upper critical dimension $d_u=8$ \cite{Biroli2007,Franz2011,Franz2012} where the fluctuations due to heterogeneous dynamics become irrelevant.  In particular, a recent computational study of hard sphere dynamics in large dimensions 
\cite{Charbonneau2012-2} tested this prediction by considering the violation of the Stokes-Einstein relation, with numerical results that seemed compatible with an absence of transport decoupling - and thus mean-field behavior - for dimensions $d \geq 8$.   These numerical observations in hard spheres prompted us to consider in detail the problem of dimensional dependence of decoupling in the East model where it is expected that the hierarchical non mean-field dynamics would be present at all dimensions \cite{Ashton2005}. 

In this work we study in detail by means of extensive numerical simulations the transport properties of the East model in dimensions $d=1$ to $d=10$.  By careful 
consideration of long-time limits and finite size effects, we argue that the upper critical dimension of the East model is larger than $d=10$, the largest dimension we study.  This would be compatible with the expectation that dynamics is actually fluctuation dominated at all dimensions.  We do so by considering the relation between structural relaxation time $\tau$ and diffusion rate $D$, which in the normal liquid state obeys the mean-field like Stokes-Einstein relation (SER), $D \sim{\tau}^{-1}$.  Departure from this relation, termed transport ``decoupling'' \cite{Ediger2000}, is a manifestation of fluctuating, non mean-field, dynamics.  
Like in previous works \cite{Swallen2003,Jung2004,Schweizer2007,Swallen2009} we characterize the breakdown of the SER in terms of a ``fractional'' SER, $D\sim{\tau}^{-1+\omega}$, with $\omega>0$ encoding the degree of violation of the standard SER.  We show that for the East model $\omega > 0$, and therefore the relevance of dynamical fluctuations, for all dimensions between $d=1$ and $d=10$. 

The paper is organized as follows: In Sec.~II, we introduce the models generalization of the East model to study their SER. 
In Sec.~III, we present our results on the upper critical dimension of the East models in various spatial dimensions. 
We carefully analyze our results by performing finite size effects in Sec.~IV. 
In Sec.~V, we investigate possible correlations between enduring kinks and various timescales. 
In Sec. VI we conclude by connecting our results to the observations in atomistic simulations of Ref.~\cite{Charbonneau2012-2}.

\section{Model and simulation details}

We study the East \cite{Jackle1991,Ritort2003} model generalised to arbitrary dimensions \cite{Ritort2003,Berthier2005,Ashton2005}, with the addition of probe particles \cite{Jung2004,Jung2005} in order to study transport dynamics. 
The East model is a two state lattice model with a dynamic constraint.
The energy function of the system is defined,
\begin{equation}
E=\sum_{i=1}^N n_i \hspace{0.5cm}  (n_i=0,1).
\end{equation}
$n_i=0$ represents an unexcited and immobile state while $n_i=1$ represents the excited state that allows motion.  There are no energetic interactions between  lattice sites and therefore the thermodynamic properties of the model are trivial.
However, there are kinetic constraints that control the dynamics of the system.
The flipping rates $k^{\pm}_{i}$ at lattice site $i$ are defined,
$k^{+}_{i}=e^{-1/T}f_{i}(\{n_{\mathbf{x}}\})$ and $k^{-}_{i}=f_{i}(\{n_{\mathbf{x}}\})$. The kinetic constraint $f_{i}(\{n_{\mathbf{x}}\})$ is a facilitation function that regulates the flipping events according to
\begin{equation}
f_{i}(\{n_{\mathbf{x}}\})=1-\mathop{\prod}^{d}_{l=1}(1-n_{\mathbf{x}_i- \mathbf{\hat{u}}_l}),
\end{equation}
where $\mathbf{\hat{u}}_l$ is the unit vector in 
the $l$-the direction of a hypercubic lattice of dimension $d$.
The kinetic constraint above allows a spin flip at a given site only if at least one of its nearest neighbours in the specified directions is in the excited state.
For one dimension, only sites to the East of an excitation can flip (and thus the name of the model); in two dimensions only sites to the North or East of an excitation, and so forth.

The scarcity of excitations in equilibrium at low temperatures makes the dynamics of the East model slow and glassy.  The model is conveniently studied numerically with continuous-time Monte Carlo algorithm and the Monte Carlo with absorbing Markov chains methods \cite{Bortz1975,Ashton2005}.
To check for finite size effects, we increase the size of the system until the physical quantities measured differ less than 1\%.
We set total simulation times to be 50$\sim$100 times the relaxation time.
We vary the temperature of the system to cover over 6 orders of magnitude in the relaxation times.
We average physical quantities over 10$\sim$$10^3$ independent trajectories.

To calculate diffusion constants for particles through a supercooled liquid, we add  probe particles to our model system, cf.~\ Refs.~\cite{Jung2004,Jung2005,Berthier2005E}. 
The probe particles occupy a site on the East model lattice, but we neglect the 
back reaction on the East model dynamics, or their mutual interaction. 
After each Monte Carlo sweep, each probe particle attempts to move to a neighboring site. To mimic the effect of jamming in a supercooled liquid, a probe particle can only move if it is on an excited site of the underlying East model, and to satisfy detailed balance, they can only move if their target site is also excitated.
We then determine the diffusion constant from the mean-square displacements of the probe particles as,
$D=\lim_{t\rightarrow\infty}\langle[\Delta \mathbf{x(t)}]^2\rangle/2dt$, where $\Delta \mathbf{x}(t)=\mathbf{x}(t)-\mathbf{x}(0)$.

\begin{figure}[h]
 \begin{center}
 \mbox{
     \hspace{-17pt}
     \subfigure{
          \includegraphics[angle=0,width=0.3\textwidth]{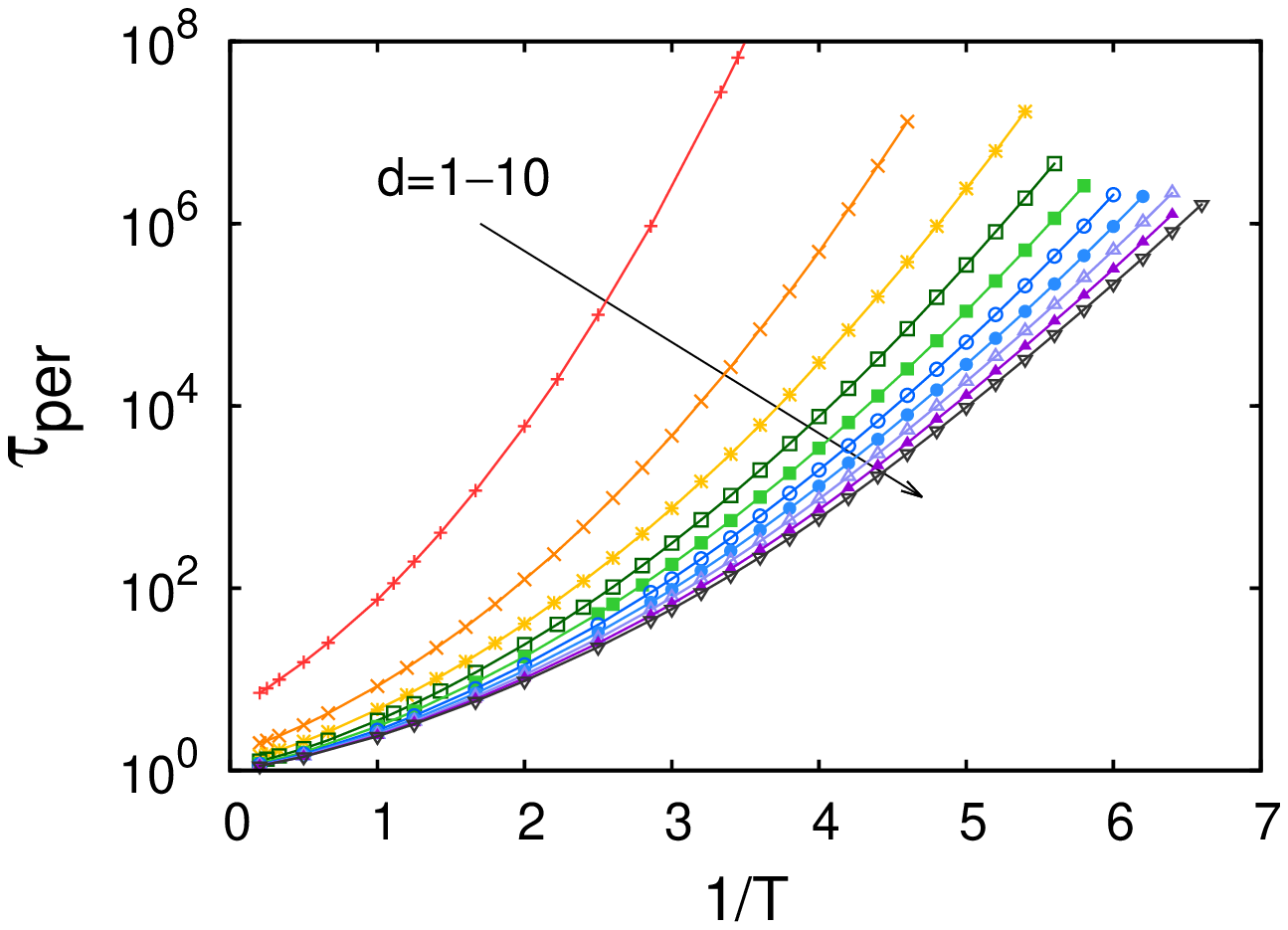}
          \label{fig1:a}
          }
     \hspace{-40pt}
     \subfigure{
          \includegraphics[angle=0,width=0.3\textwidth]{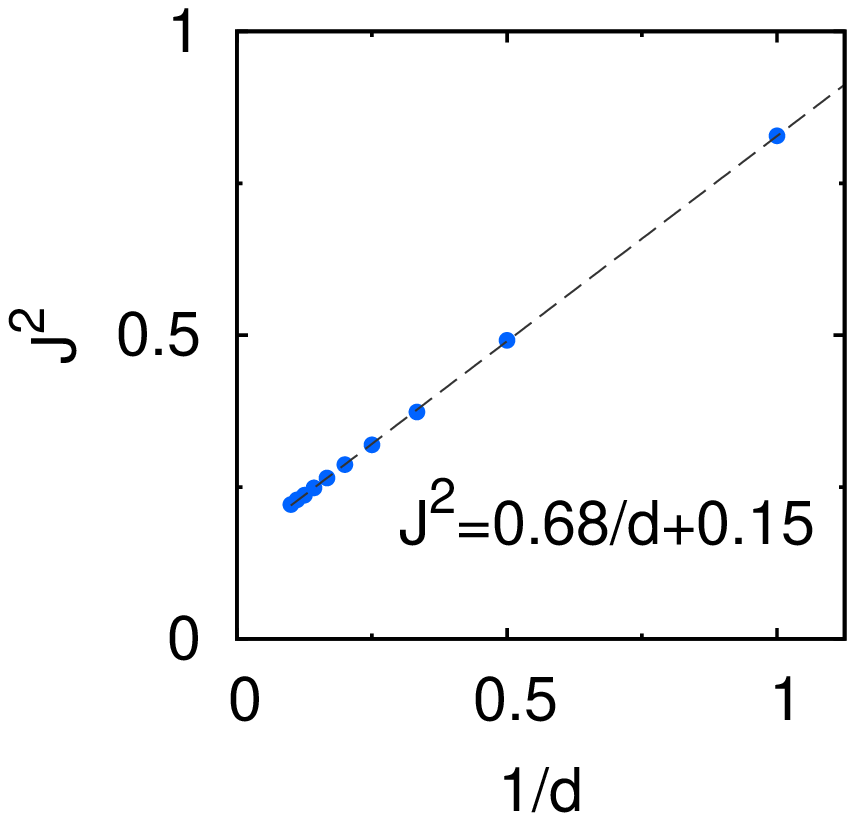}
          \label{fig1:b}
          }
 }
 \end{center}
 \vspace{-20pt}
 \caption{\label{fig1}  
(a) Temperature dependence of the mean persistence time.
In all dimensions, $\tau_{\text{per}}$ is well fitted to the Eq.~\ref{eq-tau}, which means the system is                         super-Arrhenius for all $d$.
(b) The fitting shows that $J^2$ is inversely proportional to the dimension, where $J^2$ is a fitting parameter in the Eq.~\ref{eq-tau}.
 }
\end{figure}

\begin{figure}[h]
 \begin{center}
 \mbox{
     \hspace{-17pt}
     \subfigure{
          \includegraphics[angle=0,width=\columnwidth]{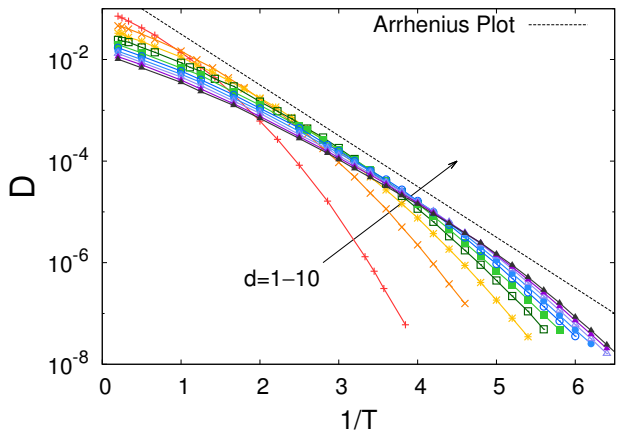}
          \label{fig2:a}
          }
 }
 \end{center}
 \vspace{-20pt}
 \caption{\label{fig2}  
                        Temperature dependence of the diffusion constants is shown. 
                            As the dimension is increased, super-Arrhenius behavior gets weaker.
 }
\end{figure}

\section{Dimensional dependence of the breakdown of the Stokes-Einstein relation in the East model}

We now investigate the properties of transport decoupling in the East model as we vary dimensionality.  If an upper critical dimension $d_u$ exists for the East model, then for $d>d_u$ the SER will be obeyed.  For this purpose we calculate the structural relaxation times and the diffusion constants for dimensions $d=1$ to $d=10$.

We use the mean persistence time of the system, ${\tau}_{\text{per}}=\langle{t_{\text{per}}}\rangle$, for the relaxation time.
The persistence time, $t_{\text{per}}$, is the waiting time at which the first flip event occurs from a randomly chosen time \cite{Jung2004}. 
The persistence time can be interpreted as the decay time of self-intermediate scattering function in the limit of large wavevector \cite{Berthier2005}. Using the mean persistence time, the relaxation time in different dimensions can be compared without wavevector dependence.

Fig. 1 shows that the mean persistence time undergoes super-Arrhenius growth for dimensions one through 10.
At fixed temperature, ${\tau}_{\text{per}}$ decreases as dimension is increased.
As expected \cite{Sollich1999,Garrahan2003,Ashton2005,Berthier2005,Chleboun2014}, the leading dependence on inverse temperature is quadratic.  In order to connect with the DF phenomenology we fit $\ln({\tau}_{\text{per}})$ with
the ``parabolic'' form \cite{Elmatad2009,Keys2011}
\begin{equation}
    \ln({\tau}_{\text{per}}/{{\tau}_{\text{o}}})={J^2}(1/T-1/T_o)^2.
    \label{eq-tau}
\end{equation}
where ${\tau}_{\text{o}}$, $J$ and $T_o$ are the fitting parameters. 
$T_o$ is the onset temperature above which the dynamics is heterogeneous
and ${\tau}_{\text{o}}$ is the relaxation time at the onset temperature.  We find that $J^2$ is inversely proportional to the spatial dimension,
$J^2 \approx 0.7/d + 0.15$.
This fit provides evidence that the dynamics in the East model is hierarchical and therefore super-Arrhenius in all dimensions.
Our fit $J^2 \propto 0.8/d$ is similar to that of Ref.~\cite{Ashton2005}.  The $1/d$ dependence we find is also consistent with the rigorous analysis of Ref.~\cite{Chleboun2014} which gives the asymptotically exact result of $J^2 = b/d$, where $b$ is $1/(2\text{log}2)\approx0.721$. 

Figure \ref{fig2} shows the corresponding numerical results for the diffusion constant $D$ of the probe particles as a function of temperature for the different dimensions.  While less pronounced than for ${\tau}_{\text{per}}$, the diffusion constant is still super-Arrhenius at all dimensions, which gets less pronounced as dimension is increased.

\begin{figure}[t]
 \begin{center}
     \hspace{-17pt}
     \subfigure{
          \includegraphics[angle=0,width=0.5\textwidth]{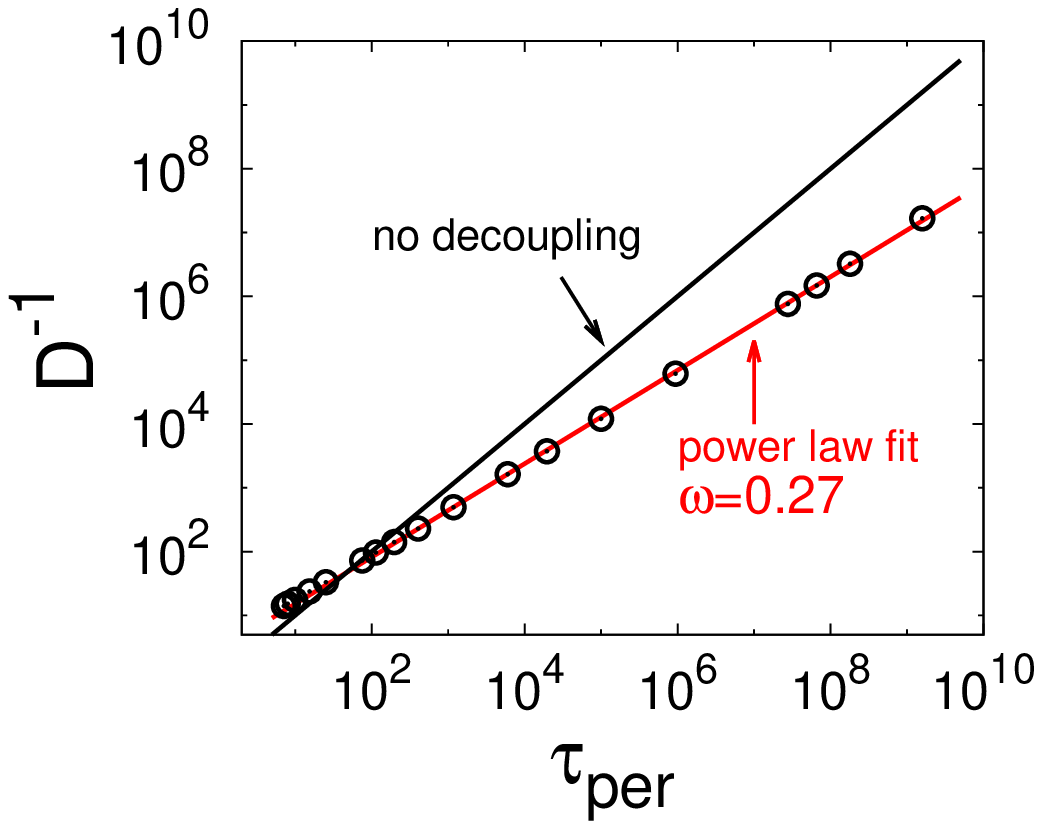}
          \label{fig3:a}
			}
     
     \subfigure{
          \includegraphics[angle=0,width=0.5\textwidth]{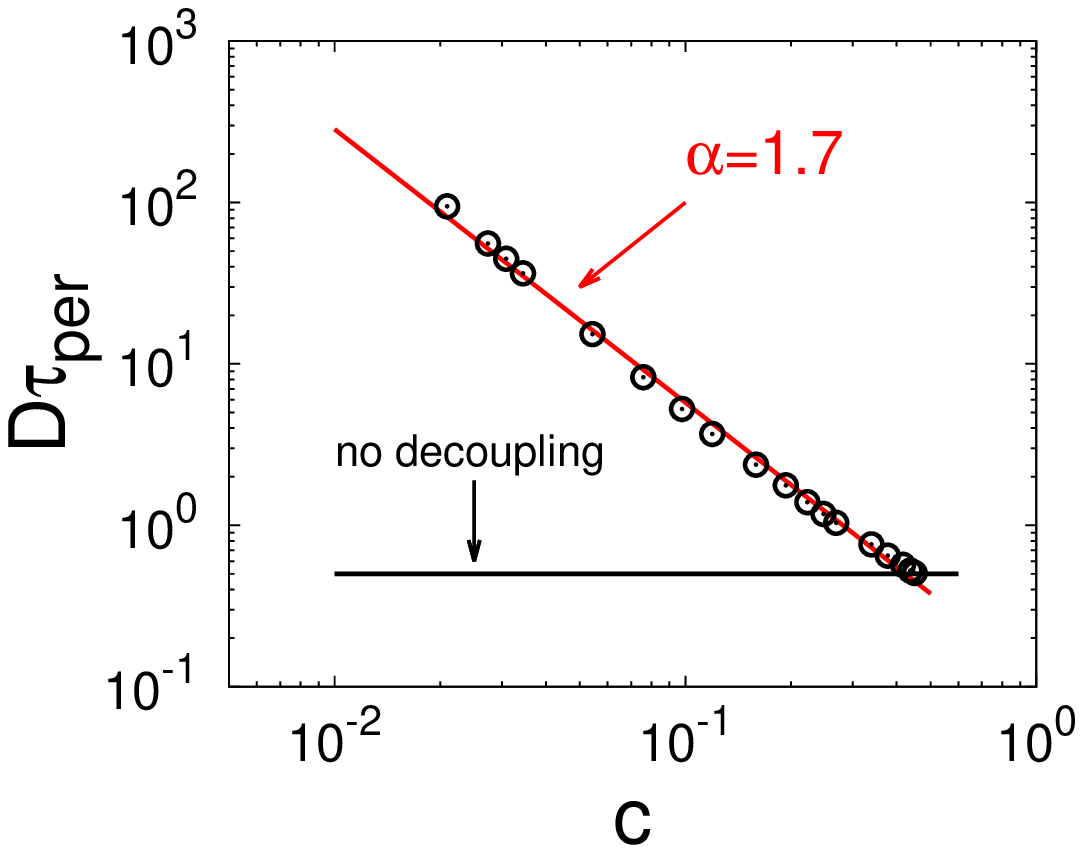}
          \label{fig3:b}
		}
 \end{center}
 \vspace{-20pt}
 \caption{\label{fig3}  
(a) Diffusion of a probe coupled to the one-dimensional East model, as in Ref.~\cite{Jung2004} (data first reported in \cite{Jung2013}). The top panel shows the inverse of the diffusion constant of the probe, $D^{-1}$, as a function of relation time $\tau_{\text{per}}$ (i.e., the mean persistence time). The red line is a fit $D\sim{\tau_{\text{per}}}^{-\xi}$ over the whole range, and the black line the case of no SE breakdown. Pronounced decoupling is obvious. (b) A test of the $D\tau_{\text{per}}\sim c^{-\alpha}$ scaling proposed in Ref., with $c=1/(1+\text{exp}(1/T))$. A best fit to the full range of data yields $\alpha \sim 1.7$, cf.~\ Ref.~\cite{Blondel2014}.
}
\end{figure}

While both the mean persistence times and the diffusion constants both show super-Arrhenius behavior, the decrease of the diffusion rate is less pronounced than the increase of the relaxation time and there is transport decoupling in the model \cite{Jung2004}.
In Ref.~\cite{Jung2004} it was originally observed that the observed decoupling in the 3 could be fit with a fractional Stokes-Einstein relation (fSER), $D\sim{\tau}^{-1+\omega}$, in analogy with the way decoupling is usually described in phenomenological observations \cite{Swallen2003}.  More recent simulations, first presented in Ref.~\cite{Jung2013}, and which we reproduce in 
Fig.~\ref{fig3:a}, extended the range of that of Ref.~\cite{Jung2004} for over nine orders of magnitude.
The range of conditions considered in Fig.~\ref{fig3:a} is the range of variation in $\tau_{\text{per}}$ that is accessible to reversible glass-forming melts.
For that range, the graphed results can be fit with a fSER,
$D \propto \tau_{\text{per}}^{-1+\omega}$ with $\omega \approx 0.27$.
The value of the exponent is consistent with those used to fit experimental data,\cite{Swallen2003}
and it is consistent with value first considered in Ref.~\cite{Jung2004}.  

Diffusion of a probe particle in the East model was also studied rigorously in Ref.~\cite{Blondel2014}.  There it was found that in the limit very low temperature the inequality $c^2 \leq D\tau \leq 1/c^{\alpha}$ holds, where $\alpha > 0$ and $c$ is the equilibrium concentration of excited sites, $c = (1+e^{1/T})^{-1}$.
While this implies breakdown of SER, it excludes fSER as $T \rightarrow 0$ because $\tau$ grows faster than any power of $1/c$ upon lowering temperature $T$.  While heuristics suggest $\alpha=2$ is suggested in the theoretical work, the best fit, shown in Fig.~\ref{fig3:b}, gives instead $\alpha \approx 1.7$.  Overall,  Figs.~\ref{fig3:a} and \ref{fig3:b} show that a fSER works extremely well as an effective description of decoupling in the relevant temperature range, and prohibitively long simulations would be required to fully clarify the scaling at vanishing temperatures \cite{Jung2013}.

\begin{figure}[t]
 \begin{center}
  \mbox{
     \hspace{-40pt}
     \subfigure{
          \includegraphics[angle=0,width=0.45\textwidth]{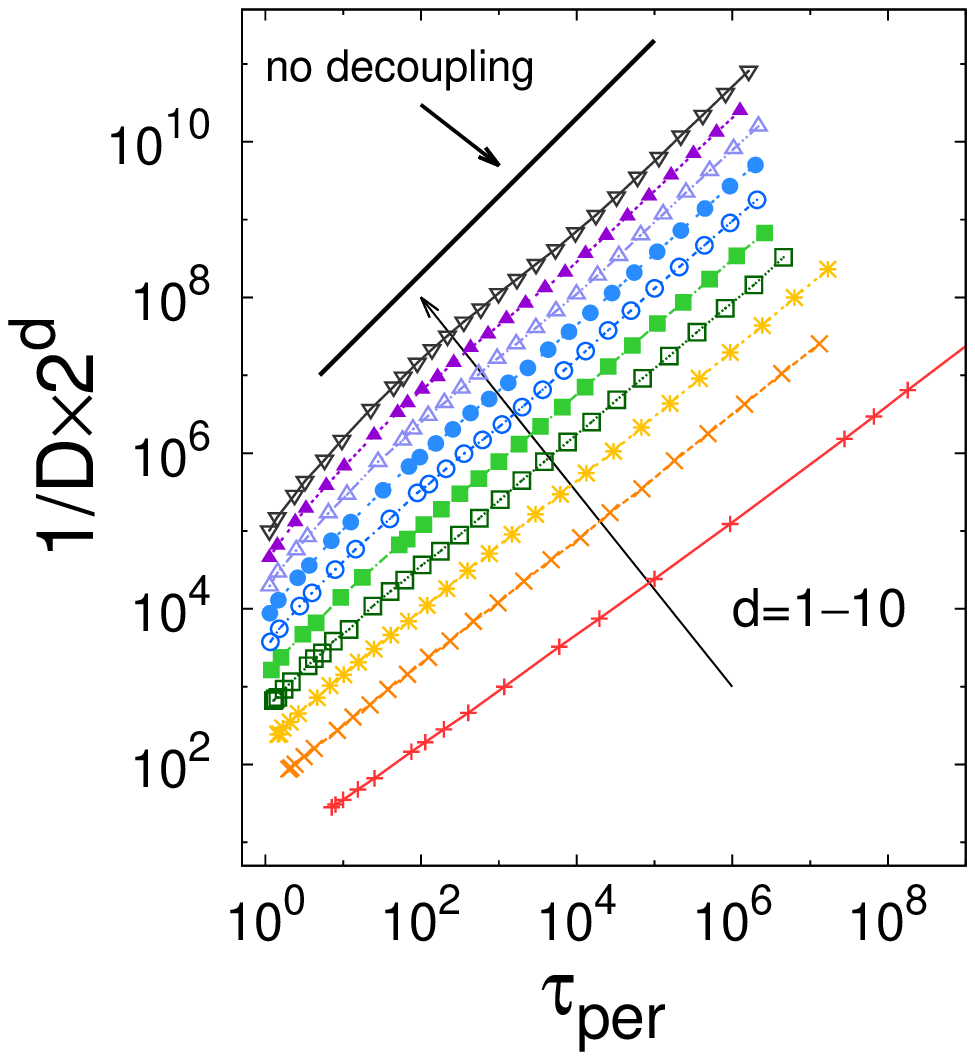}
          \label{fig4:a}
          }
     \hspace{-120pt}
     \subfigure{
          \includegraphics[angle=0,width=0.45\textwidth]{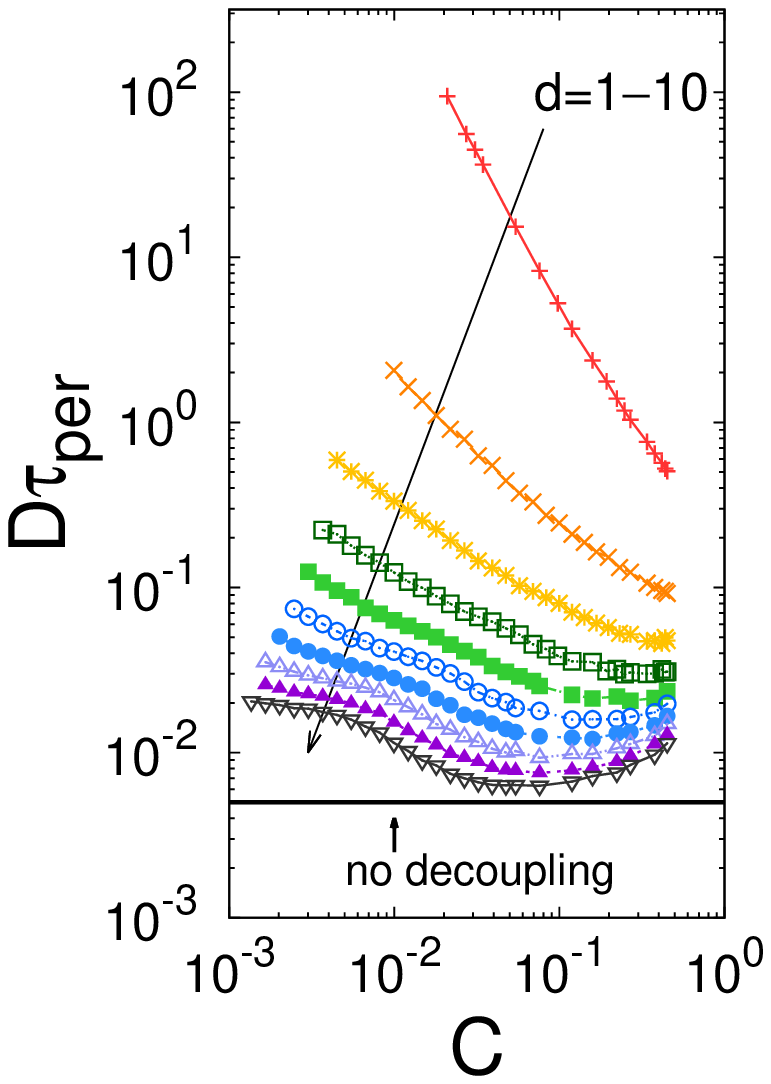}
          \label{fig4:b}
          }
 }
  
 \end{center}
     \vspace{-10pt}
 \caption{\label{fig4} (a) The power law relations between ${\tau}_{\text{per}}$ and the diffusion constant, $D\sim{\tau}^{-1+\omega}$, are shown. 
                       Data points which have ${\tau}_{\text{per}}$ longer than $10^4$ MC steps are used for the power-law fitting. Temperature
                       decreases at $\tau_\text{per}$ increases.
                      (b) The power law relations between $D\tau_\text{per}$ and the concentration of excitations, $c$, are shown. 
                       Temperature increases as $\tau_\text{per}$ increases, so the low temperature asymptotic fit is taken over concentrations below 
                       those which have $\tau_\text{per}$ longer than $10^4$ MC steps.
                        }
\end{figure}

\begin{figure}[t]
 \begin{center}
     \subfigure{
          \includegraphics[angle=0,width=\columnwidth]{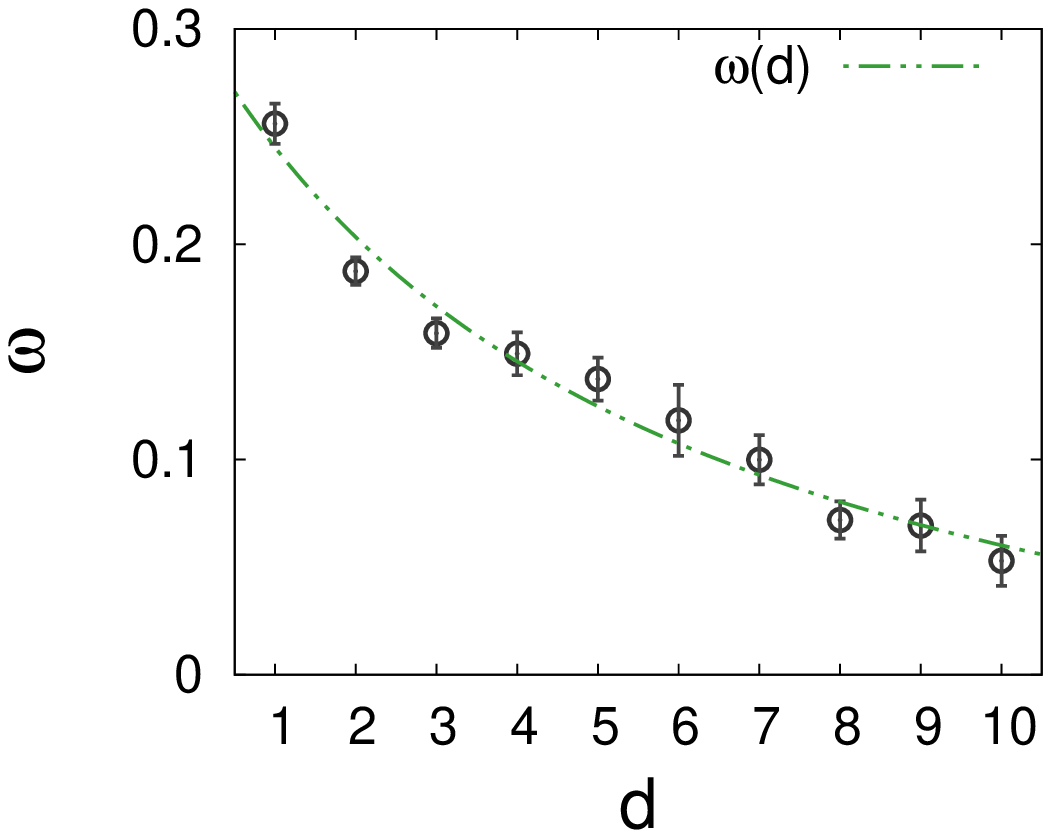}
	\label{fig5:a}
     }
     \subfigure{
          \includegraphics[angle=0,width=\columnwidth]{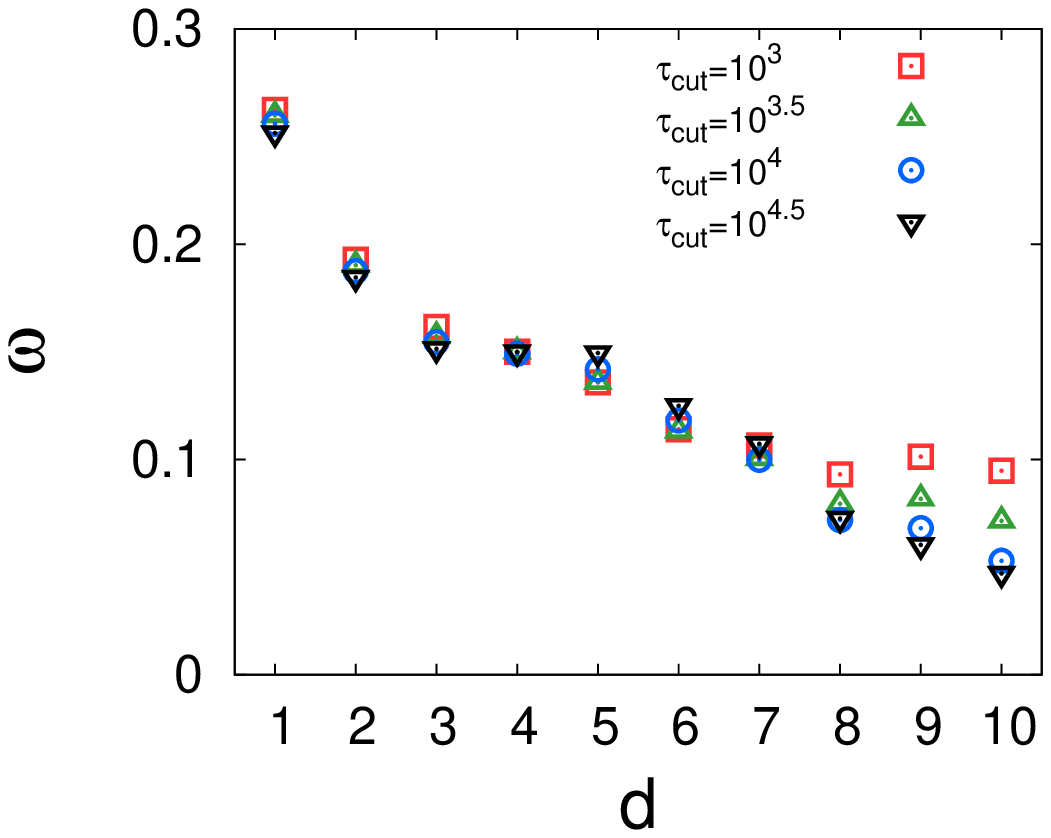}
	\label{fig5:b}
	} 
 \end{center}
 \vspace{-20pt}
 \caption{\label{fig5}  
 		         (a) The power law exponents obtained using data points in Fig.~\ref{fig4:a}.
                         $\omega$ decreases non-linearly as the dimension is increased.
                         Based on the result, the upper critical dimension, $d_u$, is not found up to $d=10$.
                         This result is distinctive from the result of the hard sphere system which shows the linear decrease and $d_u=8$.
                         We suggest the fitting form $\omega(d)=(A/d+B)/(C/d+D)$.
                         The fitted graph shows a good agreement with the data points.
                         The result shows that the exponent ${\omega}$ decreases slowly to 1
                         and supports that $d_u$ to be infinite in our model system.
                         (b) The power law exponent, $\omega$, is shown with different minimum values of $\tau_\text{per}$, which we name $\tau_\text{cut}$.
                         As higher temperature data is included ($\tau_{\text{cut}}$ is reduced), $\omega$ in $d>7$ varies significantly. 
			Using the data in Fig.~\ref{fig2}, the power exponents $\omega$ in $d>7$ converge when $\tau_{\text{cut}} >10^4$.
			The exponents in lower dimensions converge for even lower values of $\tau_\text{cut}$.
			To be consistent across dimensions, we chose $\tau_\text{cut}=10^4$ for the calcuations in Fig.~\ref{fig5:a}.  
		}
\end{figure}

Decoupling between mean persistence times and diffusion constants is also found in higher dimensions.  In Fig.~\ref{fig4} we show both $D$ against $\tau$, in order to test the validity of a fSER, and $D \tau$ as a function of $c$, to test higher dimensional versions of the asymptotic scaling of Ref.~\cite{Blondel2014}.  From both representations the presence of decoupling up to dimension $d=10$ is evident.

We obtain the fSER exponent $\omega$ by linear fitting.
The value of $\omega$ for the higher dimensions considered is sensitive to the exact fitting procedure used.  To minimize the error and to investigate the systems in the low temperature limit, we use only data where the mean persistence time is longer than $10^4$ MC sweeps. We then recalculate $\omega$ removing one data point at a time from the high temperature end of the data, stopping when we have only five data points left.
We define the error bar as half of the difference between the maximum exponent and the minimum exponent
from the varying number of data points we used.
Our results for the fSER exponent $\omega$ are shown in Fig.~\ref{fig5:a}.

At a minimum, this result demonstrates that the East model violates standard SER up through 10 dimensions. The degree of violation, $\omega$, also appears to be decaying very slowly, consistent with the hypothesis that the upper critical dimension may be infinite. The decay of $\omega$ versus $d$ does not fit well to a line. As an alternative, we consider, 
$\omega(d) = (A/d+B)/(C/d+D)$, similar to Eq.~\ref{eq-tau}.  This form fits reasonably well. Although fitting four free parameters to 10 data points is far short of a proof, it demonstrates that the data do not simply extrapolate to a finite upper critical dimension. To ensure that we have reached the long time limit in all dimensions, we vary the minimum persistence time at which we begin fitting the asymptotic slope. Fig.~\ref{fig5:b} shows that the slopes appear to have plateaued at the cutoff we have chosen, but that lower choices would have given meaningfully different results. Other systems, including hard spheres, could be subject to similar sources of error.

\begin{figure*}[t]
\includegraphics[width=\textwidth]{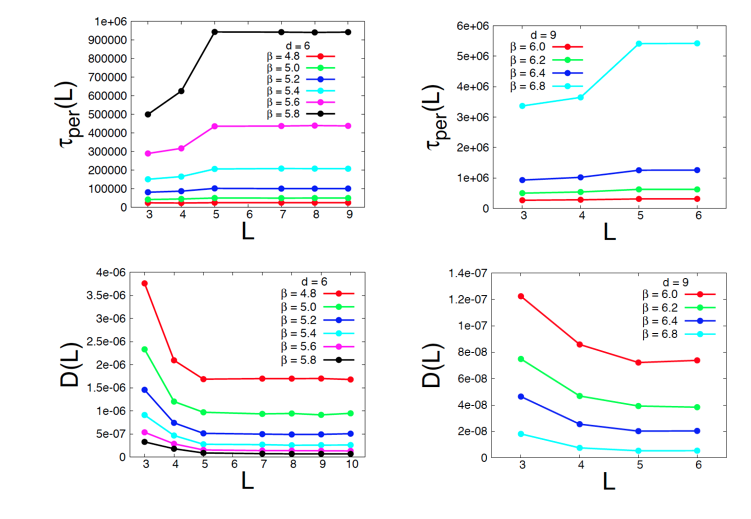}
\caption{\label{fig6} 
System size dependence on the mean persistence time (top) and of the diffusion constant (bottom) is shown in $d=6$, $d=7$, $d=8$, and $d=9$. As the system size $L$ is increased, $\tau_{\text{per}}(L)$ increases and converges to a constant value.  
$D(L)$ similarly decreases as it converges.
}
\end{figure*}

\begin{figure}[t]
 \begin{center}
\includegraphics[angle=0,width=\columnwidth]{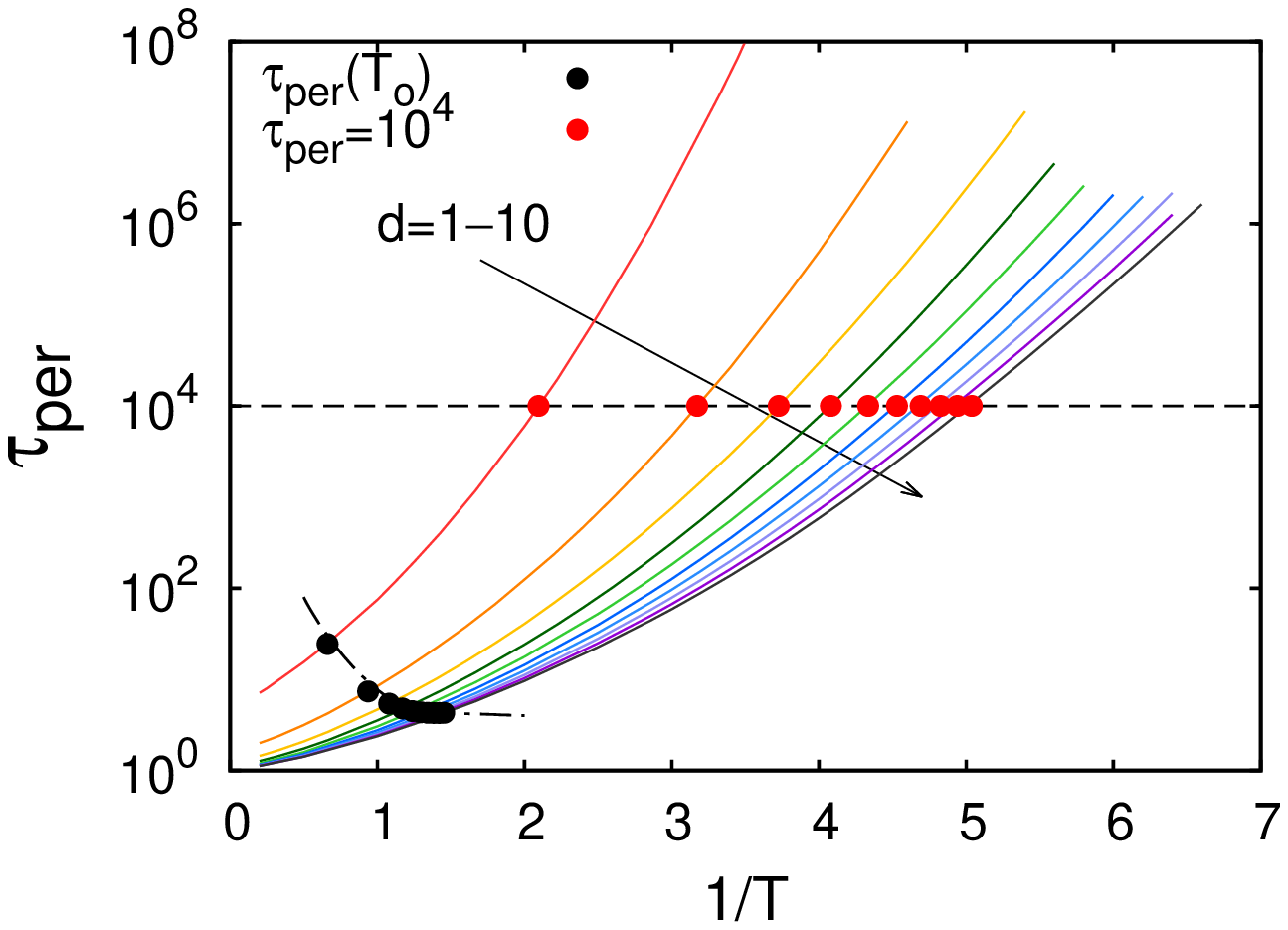}
\includegraphics[angle=0,width=\columnwidth]{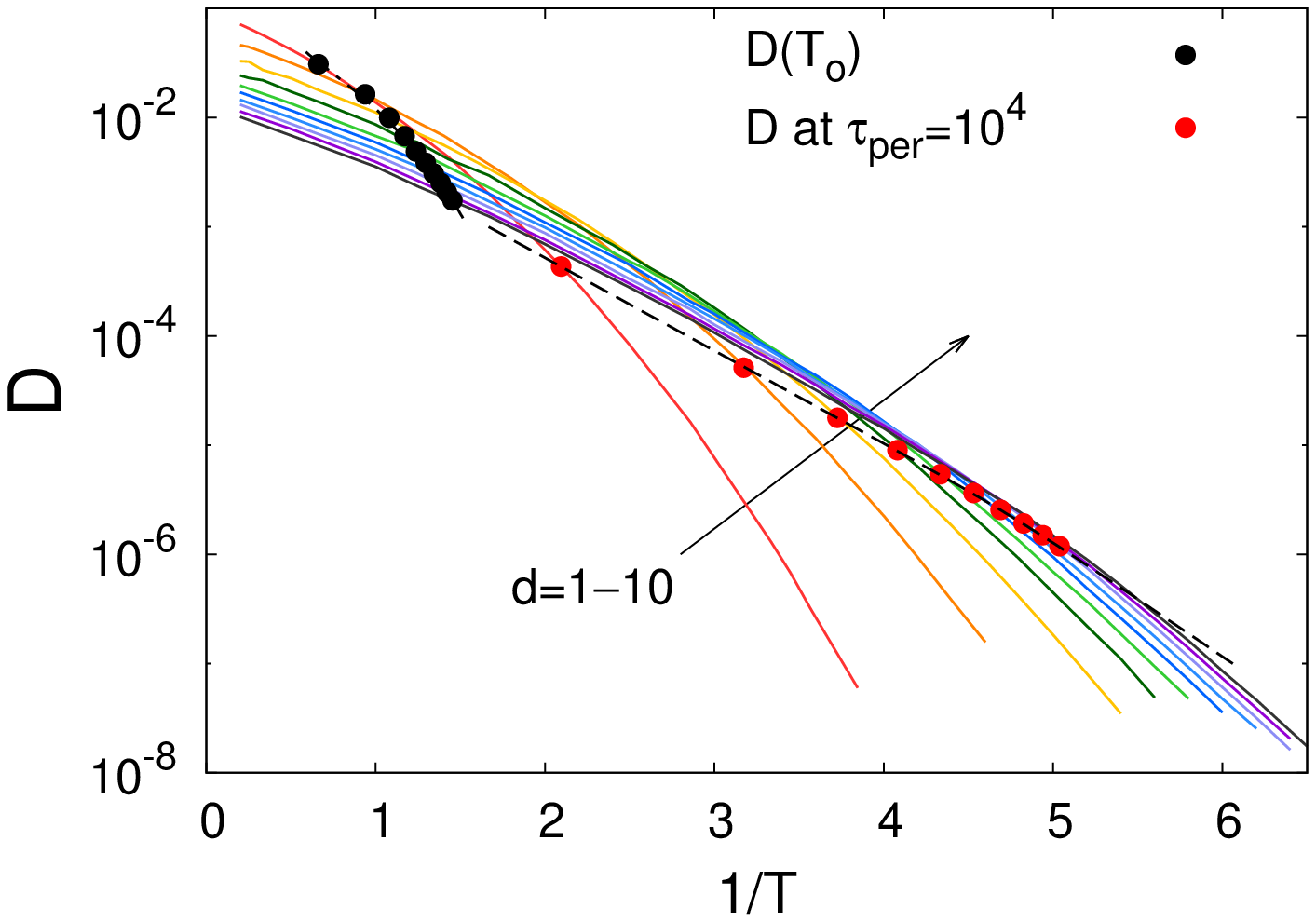}
 \end{center}
 \caption{\label{fig8} (a) Two specific time scales are marked with black dots ($\tau_{\text{per}}(T_o)$) and red dots ($\tau_{\text{per}}=10^4$).
				(b) Corresponding $D$ value is also marked.
               }

\end{figure}

\section{Finite size effects and Asymptotic behavior}

To ensure the reliability of our results in high dimensions, we check for possible finite size effects.
Fig.~\ref{fig6} shows the system size $L$ dependence of the values of $\tau_{\text{per}}$ and $D$.
In the case of $\tau_{\text{per}}$, there are no significant finite size effects when $d \leq 9$ and $L$ is near the values we used for the data already reported.
For $d=9$, the difference between $\tau_{\text{per}}(L=5)$ and $\tau_{\text{per}}(L=6)$ is less than $1\%$ for each temperature.
For $d=10$, however, the difference between $\tau_{\text{per}}(L=4)$ and $\tau_{\text{per}}(L=5)$ is more pronounced at about 30\% at the lowest temperature.
For $d=9$, the difference between $\log D(L=5)$ and $\log D(L=6)$ is less than 2\% for each temperature.
Similar to the case of $\tau_{\text{per}}$, for $d=10$, the difference is much lager and it is about 30\% at the lowest temperature.
Based on these results, our model system does not show significant finite size effects up to $d \leq 9$.
Even though $d=10$ does show stronger finite size effects at low temperatures, this does not affect the conclusion from the numerics that the upper critical dimension of the East model is greater than $d=10$, neither the slowly decreasing value of $\omega$ with dimension.

We also check that our data is in the asymptotic region compared to the onset of the heterogeneous dynamics.
To confirm whether we have reached the proper asymptotic limits in various dimensions,
we try the following variations in the fitting.
First, we can introduce an onset temperature by defining, $d\text{ln}(\tau_{\text{per}})/d(1/T)=2k_B T$ at $T=T_o$.
$T_o(d)$ is defined as the temperature at which the effective barrier to relaxation becomes order of $k_B T$.
$T_o(d)$ are marked as black dots in Fig.~\ref{fig8}.
It seems that $\tau_{\text{per}}(T_o)$ is on the order of 10-100 as we vary dimension.
It is interesting to note that although $T_o$ becomes lower with $d$, $\tau_{\text{per}}(T_o)$ gets shorter as $d$ increases.
This result comes from the fact that as the dimensionality increases, the super-Arrhenius nature of the relaxation time becomes less pronounced.
Also, we can choose our cut-off time, $\tau_{\text{cut}}$, so that only the data points $\tau_{\text{per}} \geq \tau_{\text{cut}}$
are used for the asymptotic limit fitting.
When $\tau_{\text{cut}}\gg \tau_{\text{per}}(T_o)$,
the system is in the asymptotic region and $\omega$ is not sensitive to the choice of $\tau_{\text{cut}}$,
Fig.~\ref{fig5:b}.
Note that $\tau_{\text{cut}}=10^4$ is at least 1000 times larger than $\tau(T_o)$ at every dimension considered.

\section{Discussion}

We have shown that transport decoupling occurs in the East model for all dimensions between $d=1$ and $d=10$.  This decoupling can be quantified by means of an effective fSER.  As expected, the higher the dimension the less striking fluctuation effects, which in turn manifests as decoupling becoming less pronounced.  Nevertheless it is still present at all the dimensions we simulated, which suggests that the East model has no finite upper critical dimension above which the hierarchical character of the dynamics disappears.  

Related to this weakening of the effect of fluctuations is a decrease of the onset temperature with dimensionality.  Again this is as expected: one needs to go to comparatively lower temperatures as dimension is increased to see heterogeneous dynamics.  A consequence is that one could erroneously conclude that the East model has become mean-field at some dimension by simply comparing decoupling at some fixed temperature at different dimensions, so that that temperature is in the heteorogeneous dynamics regime at lower dimension but on the homogeneous regime at higher dimension. 
Additionally, simulating sufficiently large systems is obviously quite challenging, and here we have taken great care to demonstrate the our simulation results are not hampered by finite size effects.

Out results here should also be compared to the observation of decoupling in hard spheres in high dimension of Ref.~\cite{Charbonneau2012-2}.  As in that work we find that SER breaks down, but decoupling gets attenuated as dimension increases.  In contrast to Ref.~\cite{Charbonneau2012-2} we do not see a recovery of the SER at dimension $d=8$, but decoupling in the East model persists up to $d=10$ at least.  Given the weak nature of the decoupling, it is possible that in the more challenging setting of the hard-sphere system it is difficult to distinguish weak from zero decoupling. Secondly, comparing diffusion constants across dimensions requires very careful analysis of finite size effects, such as the one we are able to do for the simpler case of the East model.  Thirdly, the onset temperature decreases (and equivalently, the onset packing fraction increases) with increasing dimension, meaning that heterogeneous dynamics may not be apparent in high dimensional simulations simply because the challenging computational nature of reaching the necessary temperatures or densities.

\section*{Acknowledgements}
The work was supported by Samsung Science and Technology Foundation under Project Number SSTF-BA1601-11.
DGT was supported by the National Science Foundation Graduate Research Fellowship under Grant No. DGE 1106400.

\bibliography{main}

\end{document}